# Insidious Imaginaries: A Critical Overview of AI Speculations

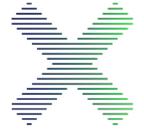

Speculative thinking about the capabilities and implications of artificial intelligence (AI) influences computer science research, drives AI industry practices, feeds academic studies of existential hazards, and stirs a global political debate. It primarily concerns predictions about the possibilities, benefits, and risks of reaching artificial general intelligence, artificial superintelligence, and technological singularity. It permeates technophilic philosophies and social movements, fuels the corporate and pundit rhetoric, and remains a potent source of inspiration for the media, popular culture, and arts. However, speculative AI is not just a discursive matter. Steeped in vagueness and brimming with unfounded assertions, manipulative claims, and extreme futuristic scenarios, it often has wide-reaching practical consequences. This paper offers a critical overview of AI speculations. In three central sections, it traces the intertwined sway of science fiction, religiosity, intellectual charlatanism, dubious academic research, suspicious entrepreneurship, and ominous sociopolitical worldviews that make AI speculations troublesome and sometimes harmful. The focus is on the field of existential risk studies and the effective altruism movement, whose ideological flux of techno-utopianism, longtermism, and transhumanism aligns with the power struggles in the AI industry to emblematize speculative AI's conceptual, methodological, ethical, and social issues. The following discussion traverses these issues within a wider context to inform the closing summary of suggestions for a more comprehensive appraisal, practical handling, and further study of the potentially impactful AI imaginaries.

## 1. Introduction

Artificial intelligence (AI) researchers and scientists have tended to make strong claims and bold predictions since their field's early days (see Mitchell 2019a, 19–21; Natale 2021, 33–49). This tendency is fuelled by the fuzziness of core theoretical assumptions (The Royal Society 2018) and by the worldviews of stakeholders acting in an exploratory domain replete with unknowns. It is also driven by opportunistic temptations to exploit the economic and sociopolitical turmoil around AI. Abiding by the entrepreneurial slogan "fake it 'till

**Dejan Grba**
Artist, researcher, and scholar, Belgrade, Serbia
dejan.grba@gmail.com

**DOI** 10.34626/2025_xcoax_001

you make it" and leveraging the cultural legacy of science fiction, the information technology and AI industries have established a thriving culture of overstating and overpromising in which speculations about AI's current, emerging, and future capabilities and consequences play a prominent part (Lorusso 2019, 64–77; Katz 2020).

Speculation is a process of conceptualizing beyond the factual. It combines imagination with the creative extrapolation of the available knowledge to predictively theorize about certain subjects (Nguyen 2021, 139). The main subjects of AI speculations are the possibilities for, and the entailing benefits and risks of, creating artificial general intelligence (AGI), developing artificial superintelligence (ASI), and reaching technological singularity. The vagueness of these concepts opens the space for unfounded claims and futuristic scenarios whose tones range from cataclysmic alarmism to deistic techno-utopianism.

The increasing application scope of AI technologies endows AI speculations with an apparition of relevance and urgency, humanity's intelligence chauvinism gives them an aura of profundity, and our love-hate relationship with technoscience supplies the emotional charge. Corporate-leaning institutes that investigate AI-related hazards aid the AI industry in proliferating the speculative AI discourse, while eager media coverage expands its reach and boosts it with sensationalism. As a result, AI has superseded nuclear technology, biotechnology, and genetics as the forefront technoscientific topic of political debates and remains among the primary sources of inspiration for films, novels, computer games, and other forms of popular culture (Marcus and Davis 2019; Mitchell 2019a 45, 215–229; 2019b).

In this paper, I draw a critical outline of speculative AI. In three central sections (2, 3, and 4), I navigate the web that interconnects AI speculations with science fiction, spiritual longings, charlatanism, dubious academic projects, questionable interests, and problematic sociopolitical worldviews. My focus (in section 4) is on the field of existential risk studies and the effective altruism movement, whose scholarly and activist work frequently fuses the AI-related phantasmagoria and mental gymnastics which support the Weltanschauung and aspirations of the tech industry, with potentially wide-reaching fallouts. Existential risk studies and effective altruism are not monolithic, and I reference authors who cast divergent critical views on their concepts, methodologies, and ethics. Nevertheless, the dominant narratives in existential risk studies and the prevailing practices of effective altruism encapsulate the interplay between philosophical pondering, techno-utopian fantasies, corporate interests, and cyberlibertarian ideologies, which constitute the AI regime. In the discussion (section 5), I juxtapose the main issues of this interplay into a wider critical perspective for understanding AI speculations. In conclusion (section 6), I make suggestions for a more comprehensive appraisal and practical handling of the possibly impactful AI imaginaries. Constructing a sound framework for these tasks is a matter of future work, and my goal with this paper is to provide a critical map that contributes to its foundation. That critical map may also be conducive to studying speculative AI in and beyond the specialist academic and techno-scientific sectors, as well as to the more exhaustive examination of the relationship between AI speculations and practices in AI art and other areas of cultural production.

## 2. AI and Science Fiction

As an important factor in romanticizing scientific research and technological development, science fiction also catalyses the penchant for AI speculations in the computer science and tech communities:

> Science fiction as a genre emerged in the era of modernity with its social upheavals and belief in technological progress. Basically, science fiction tells stories about and through fictional technology, but within the prevailing paradigm of scientific thought. Thus, the departure point of science fiction is a fictional but scientifically explainable novelty, a "novum", which establishes a new world different from the one we know. The scientific foundation, however, does not imply that the novum must be able to truly exist in the real world, but that it is cognitively imaginable within the story world. In this way, the novum enables "what if"–questions to speculate about the present and alternative futures in various, but cognitively plausible constellations. (Hermann 2021, 3)

The novum of AI has a long history in science fiction. It usually serves as a figurative, metaphorical, or allegorical device, but frequently regresses into a stereotype or cliché with a heavy anthropocentric baggage that exposes human exceptionalism (see Tromble 2020, 5).[1] The flow between science fiction and technoscience is bidirectional, so the fantasies of Arthur C. Clarke, Isaac Asimov, Ursula K. Le Guin, Philip K. Dick, William Gibson, Neal Stephenson, and many other authors have been important sources of inspiration and ideas for several generations of scientists and engineers. The influence is so strong that some college courses have been devised to employ science fiction as an effective entry point for teaching AI and computer science (Tambe et al. 2008; Burton et al. 2015). However, science fiction's inspirational appeal can also limit creative and critical thinking by "sweetening" or canonizing certain concepts regardless of their validity and instead of alternative ideas, and by stimulating aspirations that may ultimately prove meaningless or dangerous.

As a prominent subgenre, the science-fictional AI has become a reference point in a larger corpus of speculative discourse on AI's technical features, risks, and ethics (Broussard 2018; Schmitt 2021). If taken with critical caution and reserve, science-fictional AI may provide metaphors for technoscientific or philosophical thought experiments about the human condition and sociopolitical issues, but not for valid foresights or technology assessments because it is, first and foremost, a dramatic fiction tailored for popular consumption (Giuliano 2020; Hermann 2021). Like traditional narrative forms, science fiction reflects and reinforces our dispositions and behavioural tendencies with all their flaws and constraints. This may be valuable as entertainment and purposeful as a creative driver, but is insufficient for discovering new ways of thinking and knowing (Kay 1995).

A related problem of science-fictional inspiration stems from the genre's hyperproduction and acutely uneven artistic strength; there is an apparent mismatch between the popularity of science fiction works and their stylistic accomplishments, psychological inquests, or philosophical insights (see Kornbluth 1957, and Sontag 2017). The prose of many renowned science fiction authors tends to be pretentious, intellectually vacuous, operatic, verbose, self-indulgent, turgid, technophilic, gimmicky, kitschy, or schlocky (Ra 2022; Casella Brookins 2023). Regardless of its artistic merits, science fiction is criticizable as a tacit ideological platform that coerces youthful imagination and makes it productive by conjuring up enjoyable narratives about fantastic technological accomplishments and fictional worlds, which motivate young consumers to actualize them as adults (Nguyen 2021, 127–165).

Furthermore, the common affinity among scientists and engineers for science fiction's nerdy, Borgesian armchair-type adventurism may signal the insensitivity to recognizing and appreciating the emotional wonders, intellectual challenges, creative motives, and relational nuances in the real world (Kelion 2013). The capacity for this kind of recognition and appreciation is one of the contributing factors to psychological maturity and social responsibility, which are chronically iffy in the computer science and tech industry communities (Grba 2024, 13–14, 16; see also Dayan 2017; Wayne Meade et al. 2018; Jacoby 2020; and Marx 2023).

## 3. AI "Metaphysics"

The imaginaries of AGI, ASI, and technological singularity that permeate science fiction in various contexts (Hrotic 2014) also serve as devices for entertaining infantile fantasies of omnipotence and immortality, mythologizing, magical thinking, and religiosity.[2] Their allure has merged with the extrapolations of computer technologies' processing volume and speed, and the socioeconomic momentum of applied AI to constitute an important "metaphysical" layer of speculative AI.

The variety of definitions of AGI (also called human-level or strong AI) reflects the lack of robust, consensual understanding of intelligence and the closely related but even more elusive concept of consciousness in AI research (Mitchell 2019a, 42). Their baseline does not exclusively refer to human intelligence but to the artificial systems' capability for solving different problems without being trained with domain-specific data (Loukides and Lorica 2016; Sublime 2024, 42). However, AGI is commonly understood as a type of AI that matches or surpasses human problem-solving capacities, reasoning, learning, planning, and creativity across a wide range of tasks or domains (Leffer 2024).[3]

This notion accords with the computational theory of the mind (computationalism), which proposes that the mind is a computer and that mental processes are essentially computational. Influenced by Alan Turing's separation of the mind from material reality (1950), which continues the tradition of mind/body dualism reaching back at least to Descartes,[4] computationalism adopts the notion of intelligence as an unembodied mathematical construct that "operates" physical entities. It posits that the principles of universal Turing machine computation allow us to capture the essence of human thought in practice and all thought in principle, which implies that a computer programmed to reproduce human cognitive functions should be deemed intelligent (see Dietrich 1990; Browne and Swift 2019, 3; Rescorla 2020; and Natale 2021, 36). Computationalism is popular among AI scientists and some philosophers but faces strong opposing views and evidence about intelligence and consciousness as evolved biological phenomena, not substrate-independent information processors (see Damasio 1994; Steels and Kaplan 1999; Koch 2018 and 2019; Mitchell 2019b, 7; and Seth 2021).

The idea of computers that emulate human problem-solving and learning abilities often fallaciously implies their increased autonomy, leading to speculations that generally intelligent systems would quickly become superintelligent (see McQuillan 2022, 89). They would absorb the totality of human knowledge through the available learning data and combine their speed with reasoning mechanisms to make discoveries that would further increase their cognitive capacities and allow them to greatly surpass humans in all intelligence-related areas, such as scientific creativity, general wisdom, and social skills.[5] Such speculations frequently assume that generally intelligent computers will be capable of uninhibited self-improvement (both algorithmic and material), ultimately leading to the hyper-mental power and functionality of ASI, with either utopian or dystopian consequences.

The concept of runaway AI threads from computer scientist John von Neumann, who had considered but ultimately rejected it in the mid-1940s (Shanahan 2015, 233; Larson 2021, 36). Two decades later, Alan Turing's Bletchley Park fellow code-breaker Jack Good formulated the intelligence explosion model of self-improving AI that would lead to "ultraintelligence" (Larson 2021, 33–35). By the late 1980s, computer scientist Raymond Kurzweil, robotics pioneer Hans Moravec, and computer scientist/science fiction author Vernor Vinge extended these concepts into the eschatological vision of technological singularity (Vinge 1993; Kurzweil 2005; Shanahan 2015; Irrgang 2020; Larson 2021, 44–49).

Technological singularity is a hypothetical historical point at which the cumulative progress in computation, nanotechnology, genetic engineering, and other areas of science and technology will accelerate to such an extent that biological (human) and machine intelligence will merge in an exponential and irreversible but benevolent co-evolution (Mitchell 2019a, 45–50). Fusing religion with technophilia, singularity is one of the tenets of transhumanism (discussed in section 4.1) and is often derisively called "the rapture of the geeks" (Giuliano 2020, 8). However, singularitarian assumptions that the constraints of material existence can be transcended by using material technology are unsubstantiated and the beliefs that the corporate tech sector would share or renounce its accrued privileges for the benefit of all humanity look suspicious in the light of its wasteful, exploitative, and oligarchic reality (Lorusso 2019, 21–25; McQuillan 2022; Rushkoff 2022).

## 4. AI and Existential Risk Studies

Many speculations about advanced AI revolve around scenarios that connect it with existential imperilment. They figure prominently in the field of existential risk studies, which combines an offshoot of analytic philosophy, utilitarian ethics, quantification, statistical risk analysis, and imaginative thinking to explore extinction threats to humanity or all life on Earth.[6] Existential risk studies gained institutional prominence, caught public attention, and attracted the tech sector's financial backing in the first two decades of the twenty-first century. Against strong expert disagreement about the empirical basis for considering AI as an existential risk, the scholars in this field favour contemplating AI dangers over calamities that abound in factual data (Cremer and Kemp 2021, 14–15).

A prominent variety of existential risk speculations is about collateral damage that a hypothetical super-optimizer algorithm with extensive control over the material world may cause, running inconsiderately of human goals and values. Its key point is that the further development of statistical machine learning could make AI systems hard or impossible to control, not because of their intrinsic will to power but because of their simplistic reasoning mechanisms. For example, Stuart Russell's book *Human Compatible* (2019) is based primarily on extrapolating the efficiency of AI programs that optimize for fixed human-defined objectives. It suggests that, if future AI systems get "too smart too quickly", their alignment with human interests will be compromised. This concern plausibly implies that engineers may irreversibly entrust powerful AI systems with critical tasks despite their lack of commensurately strong control mechanisms such as human-level generalization and common-sense reasoning (Mitchell 2019b, 7), but also assumes that human interests are universal, neutral, and conflict-free. It is thus not surprising that Russell does not acknowledge the actual mandating trend, as absurdly reckless as the fictional one in his book, which already underpins the widespread deployment of inherently flawed and harmful AI systems (see McQuillan 2022, 27–71).

Devising a robust, broadly acceptable ethical framework is a major challenge for AI value alignment. It requires a precise, stable, and consensual common ground to build "universal" ethics for humanity. However, this necessary groundwork approach to ethics is attenuated or excised in the mainstream AI discourse mainly because of political infeasibility and technical intractability. Instead, the tech community nurtures a tacit presumption that historically, culturally, and socially divergent human interests, goals, and values are compatible and homogenous and thus can be neatly harmonized as straightforward ethical rules (see Dougherty 2001; McQuillan 2022, 35; Grba 2023; McFadden and Alvarez 2024). For instance, Max Tegmark's AI speculations in the book *Life 3.0* (2017) are remarkably pragmatic in many respects but treat ethics selectively and superficially. Tegmark mentions some obstacles to implementing basic "kindergarten ethics" into future AI but never discusses their underpinnings in the biological and historical heritage of human competitiveness or the political, economic, and cultural factors of conflicted human goals (Tegmark 2017, 283–288). He admits that interdisciplinary conversation about AI ethics is necessary because human interests and goals are not universally aligned (293) but avoids explicating that this would entail addressing present (not science-fictional) power differentials and social inequities (12–28).

A related variety of existential risk scenarios presumes that any intelligent actor will inevitably tend to dominate and exploit, based on the premise that natural selection shaped humans and other animals to use intelligence for dominance and exploitation. For example, in his book *Superintelligence* (2014), Swedish-born analytic philosopher Niklas Boström (widely known as Nick Bostrom) leverages computationalist premises to speculate about the pathways and consequences of self-modifying ASI.[7] As mentioned in section 3, the viability of computationally emulating (human) intelligence is unknown, so Bostrom's superintelligence remains a mere assumption which, like his several other works, exploits arbitrariness. There is no reason to believe that a software designed to pursue a predefined goal by optimizing some of its functions would be able to autonomously (without human input) define the concept of domination and set it as the ultimate purpose. Although Bostrom describes how an already superintelligent software might elaborate a strategy to overtake the world, he does not explain how this superintelligence would emerge in the first place. He takes the possibility of superintelligence as "obviously plausible" and therefore not requiring further explanation. By pitting human and machine intelligence in game-theoretic models of optimizing objectives, Bostrom's and Russell's thought experiments severely restrict the abilities of human cognition (Larson 2021, 34–35, 83, 279). One of the main reasons for the conceptual unevenness and ethical flimsiness in these and other AI speculations is their techno-utopian slant.

**4.1. The Techno-utopian Complex**

Bostrom's essay *Existential Risks* (2002) was one of the early works that intersected analytic and utilitarian philosophy with largely abstract extinction scenarios (Schuster and Woods 2021, 5–6). It defines existential risk as a failure to attain technological maturity (by plateauing, stagnation, or regression) or a failure to spread beyond Earth, which prevents humanity from reaching a "desirable future" and fulfilling its "long-term potential". These two types of failure may have any cause, so the definition essentially equates pressing or plausible near-term calamities, such as nuclear holocaust, with more distant hypothetical futures where humans attain sustainability and equitability but without major technological progress. Notably, Bostrom nominally identifies the subject of his existential concern as the "Earth-originating intelligent life" but favours the progressive technological enhancement of human physiological and mental capabilities (especially intelligence) toward a transhuman condition as requisite for survival and therefore a moral and ontological imperative (Schuster and Woods 2021, 16). This, and later Bostrom's publications on existential risk, drew praise in the tech community and some academic circles and were instrumental in establishing the field of existential risk studies.[8]

During the last two decades, Bostrom's philosophizing about far-off transformative technological changes, replete with science-fictional tropes and restrictive championing of intelligence, has both informed and reflected the techno-utopian approach which dominates the theorization and advisement in existential risk studies (Cremer and Kemp 2021, 3 and passim; Schuster and Woods 2021, 6–8, 23–43). The techno-utopian complex clusters mutually reinforcing notions, ideas, and attitudes in three complementary belief systems: transhumanism, total utilitarianism, and strong longtermism.

Transhumanism is a movement based on the belief that it is worth exploring the technological modifications of life, which include augmented humans, cyborgs, androids, and digital simulants. It primarily seeks to hack human biological mechanisms to extend life, expand cognition, increase welfare, and enhance well-being. Transhumanists believe that, depending on continued progress in science and technology, these improvements would synergize and become exponential in scale and speed, which would lead to achieving immortality, superintelligence, and a greater capacity for pleasure (see Pilsch 2017; and Farman 2020). Although purpose, virtue, love, and justice are not central values in their utopia of "surpassing bliss and delight" (Bostrom 2008), transhumanists argue that fictional posthuman lives are more valuable than existing humans' lives because they would be much longer, richer, and marked by more "net positive experiences". Thus, from a transhumanist perspective, efforts at addressing existential risks are not primarily about preventing the suffering and passing of presently living humans but focus on preserving the human species so it could technologically enhance itself into a posthuman one, which would be the real beneficiary (Cremer and Kemp 2021, 4–5; Schuster and Woods 2021, 23–43).

This perspective follows utilitarian reasoning, where ethical choices are driven by the so-called "expected value" (discussed in section 4.2) and the notion that potential future lives are morally equivalent to existing ones. Total utilitarianism defines moral rightness as the maximization of well-being regardless of space and time and argues for (or bets on) the future as an overwhelming moral priority, based on an assumption that the number of future lives could be unimaginably large. It is closely related to strong longtermism, which emphasizes the ethical imperative of making the choices expected to have the best effect in the distant future (Greaves and MacAskill 2001). Unlike other schools of thought that focus on evident or probable long-term issues, strong longtermism is concerned with pursuing high payoffs thousands or millions of years in the future. Exceptionally valuable potential gains justify gambling with extremely low probabilities for their actualization (see MacAskill 2019; 2022b; Centre for Effective Altruism 2024). Common in all these views is the premise about far larger populations of technologically enhanced or simulated and thus happier/better beings in the deep future.[9]

Although techno-utopian conceptual underpinnings and methodologies are neither exhaustive nor exclusive for understanding the human condition and extinction risks (Schuster and Woods 2021, 18), they have a noticeable media presence and influence a range of circuits in and beyond academia. They partly motivated the establishment of the effective altruism movement and have largely informed its practices. Effective altruism "advocates the use of evidence and reason for impartially selecting causes and calculating benefits to mobilize intellectual and material resources that provide the greatest good" (Effective Altruism 2023). This definition does not necessarily require long-term speculations, and effective altruism was initially concerned with current global problems, such as health and poverty, and helped improve the standards for organizing and evaluating charitable projects. Over time, however, the rising emphasis on humanity's long-term future within the movement turned many effective altruists into longtermists (Samuel 2022).

The lively ideological resonance of transhumanism, total utilitarianism, and strong longtermism with the Silicon Valley worldviews has gained existential risk studies and effective altruism widespread popularity among the AI safety research communities and ample financial support from tech entrepreneurs.[10] This has turned Bostrom and some other existential risk thinkers and effective altruists into major public figures and won them consulting positions with AI companies and governmental advisory bodies where their expertise contributes to decisions, policies, and strategies that may affect wide swaths of society.

**4.2. Tainted Conceits**

Existential risk scholars and effective altruists claim to strive toward scientific objectivity, but the prevailing techno-utopian inclinations introduce theoretical and practical problems in these fields. Their key issues are the ambiguous but morally loaded definitions of central terms, arbitrary categorization and incoherent assessment of future risks, non-representative ethical views, overreliance on assumptions and speculations, precipitation of certain calamities by researching them, and dangerous mitigation strategies.

Bostrom's treatment of central terms, such as "expected value", "technological maturity", "desirable future", and "humanity's long-term potential", in his discussion of existential risk (summarized in section 4.1) exemplifies the abstractness, ambiguity, subjectiveness, and ideological sway of the techno-utopian approach in existential risk studies and effective altruism. Expected value is the product of multiplying a certain outcome's value by the probability of its fulfilment. This concept may be useful in a range of theoretical contexts, but difficult to apply rigorously to the areas of deep uncertainty (with low information about probabilities) such as existential risks (Cremer and Kemp 2021, 5–6). Crucially, evaluating expected value requires a particular notion of value, which opens the gates for biased intuitions and preferences. Two other key concepts—technological maturity and humanity's long-term potential—are the ultimate unknowns. Existential risk or well-being does not need to be tied to technological maturity, and it is doubtful that scholars can reliably predict hazardous or beneficial factors of future technologies. Different plausibly desirable futures often involve conflicting trade-offs, and neither existential risk thinkers nor anyone else can define them with ultimate authority. Similarly, neat techno-utopian distinctions between existential and global catastrophic risks presume firm systemic knowledge about the functioning of society and many other complex systems on Earth, which we do not have.[11]

The community of existential risk studies is noticeably homogenous regarding income, class, ideology, age, ethnicity, gender, nationality, religion, and professional background. Most scholars are white males from wealthy Western countries working in well-endowed universities and foundations (Cremer and Kemp 2021, 8, 26). Similarly, effective altruism, longtermism, and transhumanism are most popular among the members of privileged and prosperous social groups.[12]

Their combined practices have cultivated a noticeable lack of attention for references, voices, and participants from marginalized groups and colonized societies that face oppressive and perilous conditions hic et nunc (Schuster and Woods 2021, 8–9, 20). Instead, they favour unabated technological progress as vital for achieving existential security and wellbeing, and—while recognizing the risks of some technological development paths—pay less attention to changing their political or economic drivers. They often advocate the idea of differential development toward beneficial and protective technologies, such as aligned AGI, but dismiss various options to regress, relinquish, or stop the development of certain technologies as infeasible or impossible (Cremer and Kemp 2021, 17).

Another indicative partiality is that none of the prominent existential risk thinkers discuss their field's key term "existence" with the reference to existential philosophy (existentialism), which emerged as a response to existential threats in the first half of the twentieth century: social inequalities, racism, imperialism, world wars, genocides, nuclear arms race, biodiversity depletion, etc. (Crowell 2023). Existentialism provides a more robust basis for understanding existential risks because it accommodates a range of embodied affects, lived experiences, and material and ecological relations that are not reducible to the concepts of intelligence, probabilistic reasoning, and utilitarian calculations (see Schuster and Woods 2021, 12–14). It calls for understanding and taking responsibility for one's being in the real world instead of fantasizing about immortality or omnipotence. These principles do not sit well with the techno-utopian ethos.

In contrast to existential risk scholars' assertions about striving to develop plausible scenarios grounded in scientific knowledge, their speculations are often insufficiently conclusive for making sound predictions and recommendations. They are saturated with imaginaries that frequently look superficial, particularly for ignoring the principal limits of computation and material constraints,[13] while fancying continued technological progress toward blissful transhumans in interstellar settlements or simulated realities and glossing over the violence and other unwanted side effects of (failed or successful) efforts to achieve such utopias.

All these issues have prompted scholars in humanities, science and technology studies, and some existentialist risk thinkers to criticize the lack of intellectual rigor and moral credibility in existential risk studies. They emphasize the field's radical mismatch between method and object of study, view its reliance on de facto unpredictability as pseudoscientific, and deem its social unrepresentativeness undemocratic and ethically tenuous (Cremer and Kemp 2021, 6–8, 17–18; Schuster and Woods 2021, 45; Gebru 2022; Samuel 2022; Torres 2023; Anthony 2024).[14] Bostrom has been at the centre of this critique as the field's originator and most recognizable influencer. Some philosophers and scientists dismiss his ideas as inherently unscientific or religious because they deal with imaginary and unprecedented subjects that are usually impossible to describe sufficiently or discuss rationally. For instance, Massimo Pigliucci argues (2021) that Bostrom uses hypothetical technoscientific challenges as a pretext to reboot the so-called "analytic metaphysics" or "first philosophy", which has been replaced by science for several centuries.[15] Furthermore, practical remedies that Bostrom and other existential risk thinkers propose for hypothetical calamities usually endorse techno-solutionist and totalitarian ideas that inform or reinforce troublesome political views and ethically questionable enterprises in the tech sector.

**4.3. Present Dangers**

Problems of existential risk studies and effective altruism are not purely theoretical or relevant only within academia. Even though their stakeholders belong to narrow demographics and champion relatively obscure belief systems, the ideological alignment with the tech industry has attracted economic and political support for their academic establishment and placement in the corporate and government advisory bodies. In such a position, their conceptual shortcomings, methodological flaws, and biased worldviews can precipitate catastrophes by missing certain existentially risky pathways, wasting resources on incorrect speculations, and fantasizing dangerous mitigation solutions (Cramer and Kemp 2021, 20–25). The practices of existential risk and effective altruism communities have stirred controversies and materialized some of their critics' key warnings.

The case of Bostrom's paper "Vulnerable World Hypothesis" (2019a) illustrates how speculation-based recommendations for avoiding calamities can contribute to the catastrophes they intend to escape. This paper posits that certain emerging ideas may initiate or influence technological development which may become hazardous to human civilization in the long run. Among several options to protect against them, Bostrom's comprehensive solution (including policy recommendations) proposes ubiquitous AI-governed surveillance and extreme policing. The surveillance system would involve the mandatory use of "freedom tag" devices whose multiple cameras and microphones continuously track citizens and provide their behavioural data to state-employed "freedom officers" mandated to order pre-emptive strikes by drones or the police, all under the constant watch of AI that prevents misuse. Bostrom publicly discussed this AI restaging of Orwell's *1984* without irony, self-criticism, or distancing from the fact that such totalitarian measures may inspire or aid authoritarian and malicious actors (Bostrom 2019b; see also Cremer and Kemp 2021, 12, 20–26).

Although Bostrom restrains from openly advocating for the exploitative instrumentalization of the techno-utopian paradigm, he does not seem eager to criticize it, and he has developed a chronic propensity for self-assured flirting with problematic claims. For example, as a postgraduate student at the London School of Economics in 1996, he openly expressed racist views in an Extropians listserv email, where he used the N-word, arguing that white people were more intelligent than black people. In 2023, Bostrom issued an apology for this "incident" but conspicuously failed to withdraw his central contention about race and intelligence, and instead made a partial defence of eugenics (Torres 2023). As another example, Bostrom's concept of "dysgenic pressures" in *Existential Risks* (2002) reinvigorates the eugenicist idea from the 1920s and 1930s that "less intelligent" people might outbreed "more intelligent" (Torres 2023; Ciurria 2023; Anthony 2024).

Effective altruism shares many views with existential risk studies, and some of their most respected members express, promote, or tolerate racist, xenophobic, classist, and ableist attitudes (Torres 2023). Effective altruist projects that bring theory into practice often involve ethically dubious mechanisms, and some function as self-serving schemes. The movement's affinity for the cryptocurrency market, which has a proven history of criminality and impoverishment, gained infamy with the business enterprise of Sam Bankman-Fried. This vocal effective altruist was sentenced in 2024 to 25 years in prison for financial fraud and related crimes he committed running the Futures Exchange Trading company that operated a cryptocurrency exchange and crypto hedge fund between 2019 and 2022 (Anthony 2024; RationalWiki 2024b; 2024c). Bankman-Fried's fellow effective altruist entrepreneur Benjamin Delo trailblazed this modus operandi two years earlier by perpetrating money laundering at his crypto exchange BitMEX, for which he was sentenced in 2022 to a 30-month probation and payment of a ten million USD fine (Sinclair 2022).

Intentionally or coincidentally, these communities' fixation on AI-related risks also serves questionable interests in the AI industry. For instance, it is symptomatic how the leading AI companies often raise concerns about the existential dangers of AGI and call for AI re-search non-proliferation. They frequently argue that we cannot be sure that all AI companies will be able or willing to build the aligned AGI, so only the prominent ones, with strong internal regulatory mechanisms, should be licensed to develop it (see World Science Festival 2024). Of course, big AI companies' work on advanced systems is largely secretive, which prevents meaningful external in-progress testing and oversight. The existential risk overtones in this rhetoric obscure the fact that AI engineers' explicit warnings about the absurdity and dangers of blindly building unsafe products should primarily bring into question their professional credibility and ethics, as well as their employers' motivations. Rather, it adds the aura of rationality to the frightening narratives that virtue-signal social conscience and secure support for the continuation of established AI business practices instead of critiquing and correcting them. It is part of manoeuvring toward an optimal combination of privileged position, opacity, and deregulation, which would grant top AI companies even more unaccountability and policymaking power than they already have, simultaneously increasing catastrophic risks (Narayanan and Kapoor 2024, 174).

## 5. Discussion

The fundamental obstacle to foretelling the attainment, risks, and benefits of AGI, ASI, and technological singularity is the lack of a precise and comprehensive definition of intelligence that we could use to reliably gauge the artificial intelligence "levels" and evaluate their robustness. Additionally, such breakthroughs would be unique events, so it is impossible to calibrate their predictions on comparisons with historical precedents.[16] These epistemic constraints undermine most AI speculations and AI-related existential risk studies and invalidate their policy-making legitimacy. As Melanie Mitchell pointed out (2019b): "Without more insight into the complex nature of intelligence, such speculations will remain in the realm of science fiction and cannot serve as a basis for AI policy in the real world". More generally, speculations cannot sufficiently account for the exponential complexity of the future because it unfolds through continuous interaction and coevolution of technology, economy, societal relations, and other intricate systems whose forecasting is inherently unreliable.[17]

Nevertheless, in corporate, academic, and pop-cultural discourses, the overarching claims about AI's present and phantasms about its future are usually more salient than cautious or sceptical views (see Larson 2021, 1–5, 41–42). Various motives and intentions, which include funding, promotion, marketing, and capitalization, drive some AI researchers and pundits to attenuate informed criticality and articulate AI's perspectives in science-fictional terms while championing the existing pathways in AI science and industry as vital for mitigating or solving global-scale issues en route to an automated social utopia (Verdicchio 2023). For example, according to Demis Hassabis, a CEO and co-founder of DeepMind Technologies (a subsidiary of Alphabet Inc.), AI's world-transforming scope spans from biology and medicine (protein folding, drug development, discovering the origin of life on earth) through climate change and physics (solving nuclear fusion, finding the origin of the universe, understanding the nature of reality) to metaphysics (autonomous knowledge generation and innovation) (Hassabis 2024, see also McQuillan 2022, 100).[18]

Steeped in techno-solutionism,[19] such narratives use the verb "solve" to flatten difficult issues and implicitly subjugate them to the "AI magic", as in: "If we have an AI system that can help us understand mathematics, we can solve the economy of the world" (Hasani 2024). Facing AI's extant socioeconomic problems and other urgent global risks (nuclear weapons, pandemics, neoliberal economy, climate change), they come out as disingenuous or cynical. This is exemplified by the corporate AI sector's advertisement of its vital role in solving climate change based on "overlooking" significant natural resource extraction, pollution, and carbon emissions for building and running its own infrastructure. While selling products and expertise to the oil and gas industry for optimizing fossil fuel exploitation, AI companies refuse to disclose their own energy and material resources consumption. To groom a public image of sustainability and progressive environmentalism, they purchase carbon emissions credits (Dobbe and Whittaker 2019; Dryer 2021; McQuillan 2022, 23, 52).

The veneer of scientific rigor and rationality in existential risk studies of AI and related effective altruist projects serves sly business agendas and gives the tech industry an oft-undeserved aura of respectability and authority to claim its image of visionary entrepreneurship. It allows the techno-elites to consistently favour wondrous technological remedies for both far-fetched and present existential problems over concrete social or environmental initiatives (Schuster and Woods 2021, 6–7). Analogously, strong longtermism promotes universal values of stewardship (the duty of being a "good ancestor" for posterity) which hinge on the assumption that any hypothetical risk or gain, if we imagine it as existential, justifies any amount of expenditure (Fisher 2023).

The semantic open-endedness of speculative AI terminology also sets the stage for public intellectuals and science populists to gen-erate seemingly profound but ultimately empty or meaningless narratives. For instance, here is how Joscha Bach, a Vice President of Research at the AI Foundation, opened his series of talks at the Chaos Communication Congress (Bach 2016): "Artificial Intelligence provides a conceptual framework to understand mind and universe in new ways, clearing the obstacles that hindered the progress of philosophy and psychology. Let us see how AI can help us to understand how our minds create the experience of a universe." This is an overt mystification of a disciplinarily diverse field with many pathways, disputes, contradictions, and epistemological limits (see McQuillan 2022, 39–40). Another example is the bestselling physicist Michio Kaku, who frequently misrepresents his AI-infused scenarios about the utopian future as grounded in material sciences (see, for example, Kaku 2014 and 2018). On the fatalist side of the spectrum, historian Yuval Noah Harari prognosticates that even before AI "becomes sentient" large language models will usher the end of human-dominated history (pending extinction) because these models have "hacked the operating system of human civilization [language]" (Frontiers 2023, see also Narayanan 2023).

Speculative AI narratives usually rely on technocentric worldviews, abstractions, and oversimplifications to rewrite older ideas into grandiose visions (Larson 2021, 61–63). Since the links to provable facts, insights, and predictions that give plausibility to their central hypotheses are vague or too weak in comparison with the predictive mechanisms in science and engineering, yarns about the advent of ASI and technological singularity often have strong religious overtones that exploit human imagination, fears, and hopes. In these stories, we can usually replace the word "superintelligence" or "singularity" with "magic", "miracle", or "deity" without changing their meaning (Perry 2020).

## 6. Conclusion

High-profile AI speculations centre around the fundamental issues of human existence. Thus, their appraisal should not be concerned with entertainment values or intellectual aesthetics but with evaluating the usefulness in addressing these issues and with understanding other roles they may play. In this perspective, speculative AI has proved futile, counterproductive, or dangerous. Authors of popular AI speculations tend to abuse the fabulous and fascinating nature of imaginative thinking to promote and normalize corporate interests, evangelize techno-elitist worldviews, or spread meaningless fantasies. Many share cyberlibertarian ideas informed by the zeal for technologically mediated lifestyles and future visions steeped in libertar-ian notions of freedom, social life, and economics (see Grba 2024, 15–16). They reinforce anthropocentric thinking, play along with the capitalist exploitation of AI, and champion egoistic aspirations for technocratic supremacy as the "astronomical" value.

Whether fuelled by futurism or science fiction, vacuous Gedankenexperiments or wishful thinking, tech-morality or religiosity, business or political ambitions, speculative AI is utterly frivolous but not harmless. The fancy, insincerity, hypocrisy, pretentiousness, or insidiousness lurking in magical AI thinking pollute our discursive, cognitive, economic, and political spaces already replete with undisputable problems (including the AI industry) that require immediate attention (see McQuillan 2024, 47–71, 54–57; Sublime 2024). Taking hypothetical predictions and unfounded claims uncritically, we risk allowing them to become pre-emptions that shape our reality.

When speculations become informative for widely impactful political decisions, business enterprises, and creative endeavours, their inherent unreliability urges active challenges and corrections, which are not the prerogatives of techno-elites or any other exclusive socioeconomic stratum. Such ideational constructs should become objects of socially inclusive and democratically framed interdisciplinary scrutiny, and their suitability for policymaking and susceptibility to instrumentalization must be carefully deliberated. Given all conceptual weaknesses and knowledge deficiencies in AI and other transformative technologies, this deliberation should be sceptical while remaining open to potential breakthroughs. Crucially, it must not disregard immediate, imminent, and mid-term problems in favour of gambling intellectual and material resources on dreamed-up projections of an unpredictable future.

Facing the scope and intricacy of AI science and technology and their economic and political power games, these closing notes may serve only as generic suggestions for devising more comprehensive and robust public oversight and regulatory mechanisms. Defining recommendations for a responsible and socially productive engagement with AI imaginaries, exploring speculative AI's subtler effects in academic and techno-scientific sectors, and investigating its influence on art practices, education, and other forms of cultural production requires further work.

## Notes

**1.** For instance, the plots of Isaac Asimov's novelette *The Bicentennial Man* (1976) and the film *Wings of Desire* (1987, directed by Wim Wenders) are built upon the assumption that sophisticated non-human entities would strive to become human and gladly accept all the oddities that come in the package (Asimov 1976; Wikipedia 2025).

**2.** Science fiction writer L. Ron Hubbard materialized these phantasies in a series of lucrative projects. He devised a mental therapy system called Dianetics and published it in the Astounding Science Fiction magazine in 1950. Despite the overt criticism from scientific and medical communities, Dianetics attracted a wide following, and Hubbard used it around 1952 as a basis to create Scientology, which quickly became a conglomerate of religious beliefs and often illegal business enterprises (see Behar 1991).

**3.** Noting that definitions of AGI and ASI vary (see Eliot 2024), I refer to their most common versions in this paper.

**4.** See, for instance, Ryle (2009).

**5.** For a more detailed discussion and examples of AI speculations in the computer science and AI industry, see Mitchell (2019a, 223–225; 2019b).

**6.** Existential risk studies should not be confused with the field of AI safety research, which deals with the present issues of conceptualizing, designing, and controlling safe and reliable AI systems.

**7.** Bostrom's earlier simulation argument article (2003) and latest book (2024) are also constructed around computationalist postulates.

**8.** Other influential existential risk scholars and works include Max Tegmark (2017), Sam Harris (2019), Toby Ord (2020), William MacAskill (2022a), Martin Rees, and Phil Torres. For the field's intellectual history, see Moynihan (2020).

**9.** For a further discussion of the techno-utopian complex, see Cremer and Kemp (2021, 2–6).

**10.** The Future of Humanity Institute at the University of Oxford, which Bostrom founded in 2005 and directed until its closure in 2024, enjoyed financial support from Elon Musk and Dustin Moskovitz, one of Facebook's co-founders. Musk and Moskovitz are also among the primary funders of the effective altruist Open Philanthropy Project, which has emitted 412,537,862 USD in research grants for studying AI risks between 2012 and 2024 (Open Philanthropy 2024). Skype cofounder Jaan Tallinn and Ethereum cofounder Vitalik Buterin co-founded the Future of Life Institute, a think tank that studies AI-related existential risks. Musk sits on its the advisory board at the Cambridge Centre for the Study of Existential Risk (Cremer and Kemp 2021, 1; Schuster and Woods 2021, 6; and RationalWiki 2024a).

**11.** See Cremer and Kemp (2021, 4–20) for a more detailed discussion.

**12.** It is also notable that transhumanism was initiated in the twentieth century by prominent eugenicists such as Julian Huxley (1968).

**13.** See, for example, Kauffman (2016).

**14.** These critics are joined by AI risk and safety researchers who offer more nuanced reflection. See, for example, Vallor (2024), Kasirzadeh (2025), and Gyevnar and Kasirzadeh (2025).

**15.** Analytic metaphysics attempts to arrive at conclusions about the material world by just thinking about it, which is ultimately a futile enterprise because endless hypothetical realities bear little relevance to the singular reality that we know of. Science provides empirical evidence that narrows down the possibilities from logical to physical (Pigliucci 2021).

**16.** For further discussion about the technical issues of future AI scenarios, see Verdicchio (2023) and Narayanan and Kapoor (2024, 150–178).

**17.** See Smil (2019), also Tetlock and Gardner (2015) who show that even the most successful forecasters' predictions reaching three to five years into the future default to chance (guessing).

**18.** Anthropic CEO Dario Amodei's wordy essay "Machines of Loving Grace" (2024) and OpenAI CEO Sam Altman's TED talk (2025) are two other instances of eloquent but misleading and self-serving corporate rhetoric which rephrases radical techno-utopian visions of the future AI goodness.

**19.** See Morozov (2013).

## References

**Altman, Sam.** 2025. "OpenAI's Sam Altman talks ChatGPT, AI agents and superintelligence." TED website. https://www.ted.com/talks/sam_altman_openai_s_sam_altman_talks_chatgpt_ai_agents_and_superintelligence_live_at_ted2025.

**Amodei, Dario.** 2024. "Machines of Loving Grace: How AI Could Transform the World for the Better." Dario Amodei's website. https://darioamodei.com/machines-of-loving-grace.

**Anthony, Andrew.** 2024. "'Eugenics on steroids': the Toxic and Contested Legacy of Oxford's Future of Humanity Institute." The Guardian. https://www.theguardian.com/technology/2024/apr/28/nick-bostrom-controversial-future-of-humanity-institute-closure-longtermism-affective-altruism.

**Asimov, Isaac.** 1976. "The Bicentennial Man." In *Stellar-2*. New York: Ballantine Books.

**Bach, Joscha.** 2016. "Machine Dreams - Dreaming Machines." Joscha Bach's website. December 28. http://bach.ai/machine-dreams.

**Behar, Richard.** 1991. "The Thriving Cult of Greed and Power." Time Magazine Special Report (cover story), 6 May: 50. https://www.cs.cmu.edu/~dst/Fishman/time-behar.html.

**Bostrom, Nick.** 2002. "Existential Risks: Analyzing Human Extinction Scenarios and Related Hazards." *Journal of Evolution and Technology*, 9, (1). https://www.jetpress.org/volume9/risks.html.

**Bostrom, Nick.** 2003. "Are We Living in a Computer Simulation?" *Philosophical Quarterly*, 53, (211): 243–255. https://doi.org/10.1111/1467-9213.00309.

**Bostrom, Nick.** 2008. "Letter from Utopia." Nick Bostrom's website. https://nickbostrom.com/utopia.

**Bostrom, Nick.** 2014. *Superintelligence: Paths, Dangers, Strategies*. Oxford: Oxford University Press.

**Bostrom, Nick.** 2019a. "The Vulnerable World Hypothesis." *Global Policy*, 10, (4): 455–476. https://doi.org/10.1111/1758-5899.12718.

**Bostrom, Nick.** 2019b. "How civilization could destroy itself – and 4 ways we could prevent it." TED website (00:15:18–00:19:04). https://www.ted.com/talks/nick_bostrom_how_civilization_could_destroy_itself_and_4_ways_we_could_prevent_it/transcript.

**Bostrom, Nick.** 2024. *Deep Utopia: Life and Meaning in a Solved World*. Washington: Ideapress.

**Broussard, Meredith.** 2018. *Artificial Unintelligence. How Computers Misunderstand the World*, 67–85 (71–72). Cambridge: The MIT Press.

**Browne, Kieran, and Ben Swift.** 2019. "The Other Side: Algorithm as Ritual in Artificial Intelligence." *Extended Abstracts of the 2018 CHI Conference on Human Factors in Computing Systems,* Paper No. alt11: 1–9. https://doi.org/10.1145/3170427.3188404.

**Burton, Emanuelle, Judy Goldsmith, and Nicholas Mattei.** 2015. "Teaching AI Ethics Using Science Fiction." *AAAI Workshop: AI and Ethics.*

**Casella Brookins, Jake.** 2023. "An Anti-Defense of Science Fiction." Ancillary Review of Books. https://ancillaryreviewofbooks.org/2023/12/31/an-anti-defense-of-science-fiction/.

**Centre for Effective Altruism.** 2024. "Longtermism." Centre for Effective Altruism website. https://www.centreforeffectivealtruism.org/longtermism.

**Ciurria, Mich.** 2023. "Transhumanism is Eugenics for Educated White Liberals." Biopolitical Philosophy website. https://biopoliticalphilosophy.com/2023/01/19/transhumanism-is-eugenics-for-educated-white-liberals/.

**Cremer, Carla Zoe, and Luke Kemp.** 2021. "Democratising Risk: In Search of a Methodology to Study Existential Risk." arXiv. https://doi.org/10.48550/arXiv.2201.11214.

**Crowell, Steven.** 2023. "Existentialism." The Stanford Encyclopaedia of Philosophy (Spring 2018 Edition), Edward N. Zalta, ed. https://plato.stanford.edu/archives/spr2018/entries/existentialism/.

**Damasio, Antonio.** 1994. *Descartes' Error: Emotion, Reason and the Human Brain*. New York: Grossett/Putnam.

**Dayan, Zoe.** 2017. "3 Reasons Autistic Children Excel at Computer Coding." CodeMonkey website. https://www.codemonkey.com/blog/3-reasons-autistic-children-excel-at-computer-programming.

**Dietrich, Eric.** 1990. "Computationalism." *Social Epistemology*, 4, (2): 135–154.

**Dobbe, Roel, and Meredith Whittaker.** 2019. "AI and Climate Change: How They're Connected, and What We Can Do About It." AI Now. https://ainowinstitute.org/publication/ai-and-climate-change-how-theyre-connected-and-what-we-can-do-about-it.

**Dougherty, Stephen.** 2001. "Culture in the Disk Drive: Computationalism, Memetics, and the Rise of Posthumanism." *Diacritics*, 31 (4): 85–102.

**Dryer, Theodora.** 2021. "Will Artificial Intelligence Foster or Hamper the Green New Deal?" AI Now. https://ainowinstitute.org/publication/a-digital-and-green-transition-series-will-artificial-intelligence-foster-or-hamper-the-green-new.

**Effective Altruism.** 2023. "What is Effective Altruism." Effective Altruism website. https://www.effectivealtruism.org/articles/introduction-to-effective-altruism.

**Eliot, Lance.** 2024. "Sneaky Shiftiness on the Boundaries Between AI Versus artificial general intelligence and Ultimately AI Superintelligence." Forbes. https://www.forbes.com/sites/lanceeliot/2024/12/10/sneaky-shiftiness-on-the-boundaries-between-ai-versus-agi-and-ultimately-ai-superintelligence/.

**Farman, Abou.** 2020. *On Not Dying: Secular Immortality in the Age of Technoscience*. Minneapolis: University of Minnesota Press.

**Fisher, Richard.** 2023. "What is longtermism and why do its critics think it is dangerous?" New Scientist. https://www.newscientist.com/article/mg25834382-400-what-is-longtermism-and-why-do-its-critics-think-it-is-dangerous/.

**Frontiers.** 2023. "Yuval Noah Harari: AI and the Future of Humanity | Frontiers Forum Live 2023." Frontiers Forum YouTube channel (00:06:33-00:06:50). https://youtu.be/azwt2pxn3UI.

**Gebru, Timnit.** 2022. "Effective Altruism Is Pushing a Dangerous Brand of 'AI Safety'." *Wired*, 30 November. https://www.wired.com/story/effective-altruism-artificial-intelligence-sam-bankman-fried/.

**Giuliano, Roberto Musa.** 2020. "Echoes of Myth and Magic in the Language of Artificial Intelligence." *AI & Society,* (35), 4: 1009–1024. https://doi.org/10.1007/s00146-020-00966-4.

**Grba, Dejan.** 2023. "The Transparency of Reason: Ethical Issues of AI Art." In Simon Lindgren, ed. *Handbook of Critical Studies of Artificial Intelligence*, 506. Cheltenham: Edward Elgar Publishing.

**Grba, Dejan.** 2024. "Art Notions in the Age of (Mis)anthropic AI." *Arts*, 13, (5), 137 (Artificial Intelligence and the Arts special issue, edited by Francisco Tigre Moura and Mariya Dzhimova): 1–27. https://doi.org/10.3390/arts13050137.

**Greaves, Hilary, and William MacAskill.** 2001. "The case for strong longtermism." Global Priorities Institute, University of Oxford Working Paper No. 5-2021. https://globalprioritiesinstitute.org/hilary-greaves-william-macaskill-the-case-for-strong-longtermism-2/.

**Gyevnar, Balint, and Atoosa Kasirzadeh.** 2025. "AI Safety for Everyone." arXiv. https://doi.org/10.48550/arXiv.2502.09288.

**Harris, Sam.** 2019. "Can We Avoid a Digital Apocalypse?" *Edge,* 14 June. https://www.edge.org/response-detail/26177.

**Hasani, Ramin.** 2024. "How a Worm Could Save Humanity from Bad AI." TED website. https://www.ted.com/talks/ramin_hasani_how_a_worm_could_save_humanity_from_bad_ai.

**Hassabis, Demis.** 2024. "How AI is Unlocking the Secrets of Nature and the Universe." TED website, April 2024. https://www.ted.com/talks/demis_hassabis_how_ai_is_unlocking_the_secrets_of_nature_and_the_universe.

**Hermann, Isabella.** 2021. "Artificial Intelligence in Fiction: Between Narratives and Metaphors." *AI & Society*, 1435–5655. https://doi.org/10.1007/s00146-021-01299-6.

**Hrotic, Steven.** 2014. *Religion in Science Fiction: The Evolution of an Idea and the Extinction of a Genre*. London and New York: Bloomsbury Academic.

**Huxley, Julian.** 1968. "Transhumanism." *Journal of Humanistic Psychology*, 8, (1): 73–76. https://doi.org/10.1177/002216786800800107.

**Irrgang, Daniel.** 2020. "Transhumanist Eschatology." In Bruno Latour and Peter Weibel, eds., *Critical Zones: The Science and Politics of Landing on Earth*, 282–289. Cambridge: The MIT Press.

**Jacoby, James.** 2020. "Amazon Empire: The Rise and Reign of Jeff Bezos." PBS Frontline YouTube channel. https://youtu.be/RVVfJVj5z8s.

**Kaku, Michio.** 2014. The Future of the Mind: *The Scientific Quest to Understand, Enhance, and Empower the Mind*. New York: Doubleday.

**Kaku, Michio.** 2018. *The Future of Humanity: Terraforming Mars, Interstellar Travel, Immortality, and Our Destiny Beyond Earth*. New York: Doubleday.

**Kasirzadeh, Atoosa.** 2025. "Two Types of AI Existential Risk: Decisive and Accumulative." arXiv. https://doi.org/10.48550/arXiv.2401.07836.

**Katz, Yarden.** 2020. *Artificial Whiteness: Politics and Ideology in Artificial Intelligence*, 19–89. New York: Columbia University Press.

**Kauffman, Stuart.** 2016. *Humanity in a Creative Universe*, 113–114, 242–245. New York: Oxford University Press.

**Kay, Alan.** 1995. "Powerful Ideas Need Love Too!" Written remarks to a Joint Hearing of the Science Committee and the Economic and Educational and Opportunities Committee, 12 October. http://worrydream.com/refs/Kay%20-%20Powerful%20Ideas%20Need%20Love%20Too.html.

**Kelion, Leo.** 2013. "Neal Stephenson on Tall Towers and NSA Cyber-spies." BBC News, 18 September. https://www.bbc.com/news/technology-24116925.

**Koch, Christoph.** 2018. "What Is Consciousness." *Nature,* Vol 557, 10 May: S9-S12.

**Koch, Christoph.** 2019. *The Feeling of Life Itself: Why Consciousness is Widespread but Can't be Computed*. Cambridge: The MIT Press.

**Kornbluth, Cyril M.** 1957. "The Failure of the Science Fiction Novel as Social Criticism." In Basil Davenport, ed. 1959. *The Science Fiction Novel: Imagination and Social Criticism*, 49–76. Chicago: Advent Publishers. https://sciencefiction.loa.org/biographies/pohl_failure.php.

**Kurzweil, Ray.** 2005. *The Singularity Is Near: When Humans Transcend Biology*. New York: Penguin.

**Larson, Erik J.** 2021. *The Myth of Artificial Intelligence: Why Computers Can't Think the Way We Do*. Cambridge / London: The Belknap Press of Harvard University Press.

**Leffer, Lauren.** 2024. "In the Race to Artificial General Intelligence, Where's the Finish Line?" Scientific American. https://www.scientificamerican.com/article/what-does-artificial-general-intelligence-actually-mean/.

**Lorusso, Silvio.** 2019. *Entreprecariat: Everyone is an Entrepreneur. Nobody Is Safe*. Eindhoven: Onomatopee.

**Loukides, Mike, and Ben Lorica.** 2016. "What Is Artificial Intelligence?" O'Reilly. https://www.oreilly.com/radar/what-is-artificial-intelligence/.

**MacAskill, William.** 2019. "Longtermism." Effective Altruism Forum. https://forum.effectivealtruism.org/posts/qZyshHCNkjs3TvSem/longtermism.

**MacAskill, William.** 2022a. *What We Owe the Future*. New York: Basic Books.

**MacAskill, William.** 2022b. "Longtermism." William MacAskill's website. https://www.williammacaskill.com/longtermism.

**Marcus, Gary F., and Ernest Davis.** 2019. *Rebooting AI: Building Artificial Intelligence We Can Trust*, 30. New York: Pantheon Books.

**Marx, Paris.** 2023. "Elon Musk Unmasked." Tech Won't Save Us podcast. https://youtu.be/uhGiOjTCZvo; https://youtu.be/jblBWAD_95k; https://youtu.be/JWdBhWkHm20; https://youtu.be/DF3I1PZtt-k.

**McFadden, Zari, and Lauren Alvarez.** 2024. "Performative Ethics from Within the Ivory Tower: How CS Practitioners Uphold Systems of Oppression." *Journal of Artificial Intelligence Research*, 79: 777–799. https://doi.org/10.1613/jair.1.15423.

**McQuillan, Dan.** 2022. *Resisting AI: An Anti-fascist Approach to Artificial Intelligence*. Bristol: Bristol University Press.

**Mitchell, Melanie.** 2019a. *Artificial Intelligence: A Guide for Thinking Humans*. Kindle edition. New York: Farrar, Straus and Giroux.

**Mitchell, Melanie.** 2019b. "We Shouldn't be Scared by 'Superintelligent A.I.'" The New York Times Opinion, 31 October. https://nyti.ms/3255pK2.

**Morozov, Evgeny.** 2013. *To Save Everything, Click Here: The Folly of Technological Solutionism*, 5–6. New York: PublicAffairs.

**Moynihan, Thomas.** 2020. *X-Risk: How Humanity Discovered Its Own Extinction*. Falmouth: Urbanomic.

**Narayanan, Arvind, and Sayash Kapoor.** 2024. *AI Snake Oil: What Artificial Intelligence Can Do, What It Can't, and How to Tell the Difference*. Princeton: Princeton University Press.

**Narayanan, Darshana.** 2023. "The Dangerous Populist Science of Yuval Noah Harari." *Current Affairs*, 41, (March-April). https://www.currentaffairs.org/2022/07/the-dangerous-populist-science-of-yuval-noah-harari.

**Natale, Simone.** 2021. *Deceitful Media: Artificial Intelligence and Social Life After the Turing Test*. New York: Oxford University Press.

**Nguyen, Josef.** 2021. *The Digital Is Kid Stuff: Making Creative Laborers for a Precarious Economy*. Minneapolis: University of Minnesota Press.

**Open Philanthropy.** 2024. "Grants." Open Philanthropy website. https://tinyurl.com/openphilanthropy-funding-data.

**Ord, Toby.** 2020. *The Precipice: Existential Risk and the Future of Humanity*. New York: Hatchette Books.

**Perry, Lucas.** 2020. "Steven Pinker and Stuart Russell on the Foundations, Benefits, and Possible Existential Threat of AI." Future of Life Institute podcast transcript (00:07:27–00:10:10), 15 June. https://futureoflife.org/2020/06/15/steven-pinker-and-stuart-russell-on-the-foundations-benefits-and-possible-existential-risk-of-ai/.

**Pigliucci, Massimo.** 2021. "Bad Ideas in Practical Philosophy, Nick Bostrom Edition." Medium. https://figsinwinter.medium.com/bad-ideas-in-practical-philosophy-nick-bostrom-edition-e06bf9af2aca.

**Pilsch, Andrew.** 2017. T*ranshumanism: Evolutionary Futurism and the Human Technologies of Utopia*. Minneapolis: University of Minnesota Press.

**Ra, Carla.** 2022. "The Problem with Sci-fi." Carla Ra's website. https://www.authorcarlara.com/post/the-problem-with-sci-fi.

**RationalWiki.** 2024a. "Effective Altruism." RationalWiki article. https://rationalwiki.org/wiki/Effective_altruism.

**RationalWiki.** 2024b. "Where 'Effective Altruists' actually send their money." RationalWiki article. https://rationalwiki.org/wiki/Effective_altruism#Where_.22Effective_Altruists.22_actually_send_their_money.

**RationalWiki.** 2024c. "Sam Bankman-Fried." RationalWiki article. https://rationalwiki.org/wiki/Effective_altruism#Sam_Bankman-Fried.

**Rescorla, Michael.** 2020. "The Computational Theory of Mind." In Edward N. Zalta, ed. *The Stanford Encyclopaedia of Philosophy* (Fall 2020 Edition). https://plato.stanford.edu/archives/fall2020/entries/computational-mind.

**Rushkoff, Douglas.** 2022. *Survival of the Richest: Escape Fantasies of the Tech Billionaires*, 12, 79–80. New York: W. W. Norton & Company.

**Russell, Stuart J.** 2019. *Human Compatible: Artificial Intelligence and the Problem of Control*. New York: Viking.

**Ryle, Gilbert.** 2009. *The Concept of Mind*, 1–13. London and New York: Routledge.

**Samuel, Sigal.** 2022. "Effective altruism's most controversial idea." Vox. https://vox.com/future-perfect/23298870/effective-altruism-longtermism-will-macaskill-future.

**Schmitt, Philipp.** 2021. "Blueprints of Intelligence." *Noema Magazine*, 2 March. https://www.noemamag.com/blueprints-of-intelligence.

**Schuster, Joshua, and Derek Woods.** 2021. *Calamity Theory: Three Critiques of Existential Risk*. Minneapolis / London: University of Minnesota Press.

**Seth, Anil.** 2021. B*eing You: A New Science of Consciousness*, 233–252. London: Faber & Faber.

**Shanahan, Murray.** 2015. *The Technological Singularity*. Cambridge: The MIT Press.

**Sinclair, Sebastian.** 2022. "BitMEX Co-Founder Delo Gets 30 Months Probation, Avoids Jail Time." Blockworks. https://blockworks.co/news/bitmex-co-founder-delo-gets-30-months-probation-avoids-jail-time.

**Smil, Vaclav.** 2019. *Growth: From Microorganisms to Megacities*, vii–xxv, 1–69, 303–448, 509–513. Cambridge: The MIT Press.

**Sontag, Susan.** 2017. "The Imagination of Disaster." In Rob Latham, ed. 2017. *Science Fiction Criticism: An Anthology of Essential Writings*, 189–199. New York: Bloomsbury Academic.

**Steels, Luc, and Frederic Kaplan.** 1999. "Bootstrapping Grounded Word Semantics." In Ted Briscoe, ed. *Linguistic Evolution Through Language Acquisition: Formal and Computational Models*. Cambridge: Cambridge University Press.

**Sublime, Jérémie.** 2024. "The AI Race: Why Current Neural Network-based Architectures are a Poor Basis for Artificial General Intelligence." *Journal of Artificial Intelligence Research*, 79: 41–67. https://doi.org/10.1613/jair.1.15315.

**Tambe, Milind, Anne Balsamo, and Emma Bowring.** 2008. "Using Science Fiction in Teaching Artificial Intelligence." *Using AI to Motivate Greater Participation in Computer Science, Papers from the 2008 AAAI Spring Symposium, Technical Report SS-08–08*, Stanford.

**Tegmark, Max.** 2017. *Life 3.0: Being Human in the Age of Artificial Intelligence*. New York: Alfred A. Knopf.

**Tetlock, Phillip E., and Dan Gardner.** 2015. *Superforecasting: The Art and Science of Prediction*, 13, 241. Kindle edition. New York: Crown Publishers.

**The Royal Society.** 2018. "Portrayals and perceptions of AI and why they matter." The AI narratives research project findings. London: The Royal Society.

**Torres, Émile P.** 2023. "Nick Bostrom, Longtermism, and the Eternal Return of Eugenics." Truthdig. https://www.truthdig.com/articles/nick-bostrom-longtermism-and-the-eternal-return-of-eugenics-2/.

**Tromble, Meredith.** 2020. "Ask Not What A.I. Can Do for Art… but What Art Can Do for A.I." In Andrés Burbano and Ruth West, eds. "AI, Arts & Design: Questioning Learning Machines." *Artnodes*, 26: 1–9. https://doi.org/10.7238/a.v0i26.3368.

**Turing, Alan M.** 1950. "Computing Machinery and Intelligence." *Mind*, 59, (236): 433–460.

**Vallor, Shannon.** 2024. *The AI Mirror: How to Reclaim Our Humanity in an Age of Machine Thinking*. Oxford: Oxford University Press.

**Verdicchio, Mario.** 2023. "Marking the Lines of Artificial Intelligence." In Simon Lindgren, ed. *Handbook of Critical Studies of Artificial Intelligence*, 245–253. Cheltenham: Edward Elgar Publishing.

**Vinge, Vernor.** 1993. "The Coming Technological Singularity: How to Survive in the Post-Human Era". *NASA Conference Publication*, 10129: 11–22. https://ntrs.nasa.gov/api/citations/19940022855/downloads/19940022855.pdf.

**Wayne Meade, Whitney, Letha Etzkorn, and Huaming Zhang.** 2018. "The Missing Element: A Discussion of Autism Spectrum Disorders in Computer Science." *2018 ASEE Southeastern Section Conference*. American Society for Engineering Education.

**Wikipedia.** 2025. "Wings of Desire." Wikipedia article. https://en.wikipedia.org/wiki/Wings_of_Desire.

**World Science Festival.** 2024. "AI and Quantum Computing: Glimpsing the Near Future." World Science Festival YouTube channel (00:50:05–00:53:26). https://youtu.be/gZZan4JMwk4?si=TmGN70Okbd_Edgja.